# Phonon excitation and energy redistribution in phonon space for energy transport in lattice structure with nonlinear dispersion

Zhijie Xu[1,a]

1. Fundamental and Computational Sciences Directorate, Pacific Northwest National Laboratory, Richland, WA 99352, USA

## Abstract

We first propose fundamental solutions of wave propagation in dispersive chain subject to a localized initial perturbation in the displacement. Analytical solutions are obtained for both second order nonlinear dispersive chain and homogenous harmonic chain using stationary phase approximation. Solution was also compared with numerical results from molecular dynamics (MD) simulations. Locally dominant phonon modes (*k*-space) are introduced based on these solutions. These locally defined spatially and temporally varying phonon modes *k(x,t)* are critical to the concept of the local thermodynamic equilibrium (LTE). Wave propagation accompanying with the nonequilibrium dynamics leads to the excitation of these locally defined phonon modes. It was found that the system energy is gradually redistributed among these excited phonons modes (*k*-space). This redistribution process is only possible with nonlinear dispersion and requires a finite amount of time to achieve a steady state distribution depending on the spatial distribution (or frequency content) of the initial perturbation and the dispersion relation. Sharper and more concentrated perturbation leads to a faster redistribution. This energy redistribution generates localized phonons with various frequencies that can be important for phonon-phonon interaction and energy dissipation in nonlinear systems. Ballistic type of heat transport along the harmonic chain reveals that at any given position, the lowest mode ($k$=0) is excited first and gradually expanding to the highest mode ($k_{\max}(x,t)$), where $k_{\max}(x,t)$ can only asymptotically approach the maximum mode $k^B$ of the first Brillouin zone ( $k_{\max}(x,t) \to k^B$ ). No energy distributed into modes with $k_{\max}(x,t) < k < k^B$ demonstrates that the local thermodynamic equilibrium cannot be established in harmonic chain. Energy is shown to be uniformly distributed in all available phonon modes $k \le k_{\max}(x,t)$ at any position with heat transfer along the harmonic chain. The energy flux along the chain is shown to be a constant with time and proportional to the sound speed (ballistic transport). Comparison with the Fourier's law leads to a time-dependent thermal conductivity that diverges with time.

[a] Electronic mail: zhijie.xu@pnnl.gov, zhijiexu@hotmail.com, Tel: 509-372-4885

## 1. Introduction

Nonequilibrium statistical mechanics is concerned with why and how systems evolve to appropriate equilibrium state, which still is far more tentative and not entirely satisfactory [1]. A fundamental assumption made to the standard treatment of nonequilibrium steady states is a rather quick establishment of local thermodynamic equilibrium (LTE) [2], where thermodynamic quantities temperature, pressure can be defined locally and evolution equations can be written to describe the temporal evolution of these quantities. Heat conduction is a fundamental example of nonequilibrium steady states (NSS) where the driving force on the system is the temperature gradient. The LTE assumption enables us to define the temperature field $T(\boldsymbol{x})$ locally that is varying slowly in space. By this mean, the macroscopic phenomenological Fourier's law can be established as $J = -\kappa \nabla T$ from the linear response theory, where $J$ is the heat flux and $\kappa$ is the thermal conductivity. However, the question of validity of local thermodynamic equilibrium assumption and how LTE is established has not been answered so far.

Following the standard linear response theory [3-6], the thermal conductivity can be calculated by the well-known Green-Kubo formula,

$$\kappa = \lim_{t\to\infty}\left(\lim_{N\to\infty}\frac{NV_a}{k_B T^2}\int_0^t C(\tau)\,d\tau\right) \qquad (1)$$

where $C(\tau) = \langle J(\tau) J(0)\rangle$ is the autocorrelation function of heat flux $J$, $k_B$ is the Boltzmann constant, $N$ is the system size, and $V_a$ is the atomic volume. A finite transport coefficient $\kappa$ is explicitly related to the equilibrium correlations of the system and converges with increasing time $t$ and system size $N$.

The microscopic models of energy transport and heat conduction in low-dimensional systems of interacting particles has received a lot of attentions for many years [7-14]. The anomalous conductivity $\kappa \to \infty$ was first reported for the simplest model of heat conduction along one-dimensional homogeneous harmonic chain subject to temperature gradient applied through the fixed temperatures at the two opposite ends of the chain [7, 8]. It was found that the constant heat flux across the chain with increasing system size *N* violates the Fourier's law and leads to an infinity conductivity $\kappa$.

It is now generally believed that the thermal conductivity $\kappa \propto N^{\alpha}$ with increasing system size *N*, where the power $\alpha$ can be dependent on the boundary conditions and also the spectral properties of the baths [15]. It was shown for heterogeneous disordered harmonic chain the size-dependent parameter $\alpha = 1/2$ [9] for fixed end boundary and $\alpha = -1/2$ for free end boundary [11]. Some other models with nonlinear particle interactions (Fermi-Pasta-Ulam chain with quartic interaction [16]) also do not show the size-independent conductivity ($\alpha = 0$ according to Fourier's Law). Recent study found  the expected converged size-independent conductivity for heat conduction along harmonic chain but with asymmetric potential [13]. All models mentioned above focus on the determination of the dependence of conductivity $\kappa$ on the system size *N*. However, the time-dependence of $\kappa$ has not been investigated much so far.

This paper will address these questions in an analytical way and is organized as follows: Section 2 presents the elemental solution of wave propagation in a one-dimensional dispersive chain subject to an initially localized perturbation; Section 3 describes the asymptotic solutions obtained based on the stationary phase approximation, followed by the concept of locally dominant phonon modes (*k*-space); The energy redistribution in *k*-space derived based on these solutions is presented in Section 4; Solutions to the one-dimensional harmonic chains are presented in Section 5 for energy distribution in *x*-space and *k*-space. Finally, the heat transport along the harmonic chain is studied in Section 6 where the establishment of local thermodynamic equilibrium (LTE) is discussed and equivalent thermal conductivity is shown to diverge with time.

## 2. Wave propagation and energy redistribution in dispersive chain

We first consider the solution of wave propagation in an one-dimensional dispersive chain of infinite length and subject to an initial displacement perturbation that is represented by a Gaussian wave packet at time $t = 0$,

$$u(x, t = 0) = Ae^{-x^2/4(\Delta x_0)^2}, \quad (2)$$

where $u(x,t)$ is the displacement field. The initial disturbance is centered around $x = 0$, and has an amplitude of *A*. $\Delta x_0$ is the root-mean-square spatial spread of the energy distribution in the wave packet. The energy density is proportional to $|u(x)|^2$. The total energy $\Pi$ in the system is finite and can be written as,

$$\Pi = \int_{-\infty}^{\infty} B_1 |u(x)|^2 dx = B_1 \sqrt{2\pi} A^2 \Delta x_0, \quad (3)$$

where the normalization constant $B_1 = \Pi / \left(\sqrt{2\pi} A^2 \Delta x_0\right)$ has the unit of energy density. This isolated system has no external exchange of energy and mass, i.e. the total energy should be conserved. The Fourier transformation of the initial displacement field is

$$F(q, t = 0) = \int_{-\infty}^{\infty} u(x, t = 0) e^{-iqx} dx = \sqrt{4\pi} \Delta x_0 A e^{-(\Delta x_0)^2 q^2}, \quad (4)$$

where *q* is the wavenumber. The energy distribution in *q*-space can be written as

$$E(q, t) = B_2 |F(q)|^2 = \Pi \sqrt{\frac{2}{\pi}} \Delta x_0 e^{-2\Delta x_0^2 q^2}, \quad (5)$$

where $B_2 = \Pi \Big/ \left( \sqrt{8\pi^3} A^2 \Delta x_0 \right)$ satisfying the normalization condition where total energy $\Pi = \int_{-\infty}^{\infty} B_2 \left| F(q) \right|^2 dq$.

At $t > 0$ this perturbed nonequilibrium system starts to evolve with wave propagating through the dispersive chain that is characterized by a nonlinear dispersion relation up to the second order,

$$\omega = c_1 q + c_2 q^2 , \tag{6}$$

where $q$ is the wavenumber and $\omega$ is the frequency. Coefficient $c_1$ has the unit of velocity and $c_2$ is the second order dispersion coefficient that has a unit of diffusion coefficient. It is well known that the wave will not propagate with a constant shape in dispersive medium. The governing Partial Differential Equations (PDEs) to describe the dynamics can be found based on our previous study [17-20]:

$$u_{,t} + c_1 u_{,x} = i c_2 u_{,xx} , \tag{7}$$

or

$$u_{,tt} = c_1^2 u_{,xx} - 2 i c_1 c_2 u_{,xxx} - c_2^2 u_{,xxxx} . \tag{8}$$

The temporal evolution of the displacement field $u(x,t)$ can be obtained by solving Eqs. (7) or (8), or using an inverse Fourier transformation to find complex solution $f(x,t)$, where

$$f(x,t) = \frac{1}{2\pi} \int_{-\infty}^{\infty} F(q) e^{i(qx - \omega t)} dq = \frac{\Delta x_0 A}{\sqrt{\pi}} \int_{-\infty}^{\infty} e^{-(\Delta x_0)^2 q^2} e^{i(qx - \omega t)} dq . \tag{9}$$

Integration of Eq. (9) with the dispersion relation (6) leads to the solution of $f(x,t)$,

$$f(x,t) = \frac{A \Delta x_0}{\Delta x} e^{-(x - c_1 t)^2 / 4(\Delta x)^2} , \tag{10}$$

where $(\Delta x)^2 = (\Delta x_0)^2 + ic_2 t$, and $i$ is the imaginary number. Further simplification leads to the expression,

$$f(x,t) = \frac{A}{\left(1+\alpha^2\right)^{1/4}} \exp\left(-\frac{(x-c_1 t)^2}{4\Delta x_0^2\left(1+\alpha^2\right)}\right) \exp\left[i\left(-\frac{D_2}{8}(x-c_1 t)^2\right)\right] e^{i\varphi}, \tag{11}$$

where a dimensionless constant $\alpha = c_2 t / \Delta x_0^2$ is introduced to represent the effect of nonlinear dispersion. The physical displacement field $u(x,t)$ can be represented by the real part of $f(x,t)$,

$$u(x,t) = real(f) = \frac{A}{\left(1+\alpha^2\right)^{1/4}} \exp\left(-\frac{(x-c_1 t)^2}{4\Delta x_0^2\left(1+\alpha^2\right)}\right) \cos\left[-\frac{D_2}{8}(x-c_1 t)^2 + \varphi\right], \tag{12}$$

where $D_2 = \dfrac{-2\alpha}{\Delta x_0{}^2\left(1+\alpha^2\right)}$. (13)

The phase angle $\varphi$ satisfies $\tan(\varphi) = D_5 / D_4$ with asymptotic limit $\lim\limits_{t\to\infty}\varphi = -\pi/4$. All other relevant parameters used in the formulation are

$$D_1 = \frac{1}{\Delta x_0{}^2\left(1+\alpha^2\right)}, \tag{14}$$

$$D_3 = \sqrt{\frac{D_2}{2} + \frac{1}{2}\sqrt{D_2^2 + 4D_1^2}}, \tag{15}$$

$$D_4 = \frac{1}{2}\left(D_3 + \sqrt{\frac{\sqrt{D_2^2 + 4D_1^2}}{2} - \frac{D_2}{2}}\right), \tag{16}$$

$$D_5 = \frac{1}{2}\left(D_3 - \sqrt{\frac{\sqrt{D_2^2 + 4D_1^2}}{2} - \frac{D_2}{2}}\right). \tag{17}$$

A plot of displacement $u(x,t)$ is presented in Fig. 1 with $A = \Delta x_0 = c_1 = c_2 = 1$ showing decreasing amplitude and increasing spatial spread. The modulus of $f(x,t)$ can be found as,

$$\left|f\left(x,t\right)\right| = \frac{A}{\left(1+\alpha^2\right)^{1/4}} \exp\left(-\frac{\left(x-c_1 t\right)^2}{4\Delta x_0^2\left(1+\alpha^2\right)}\right). \tag{18}$$

The energy distribution in *x*-space at any time *t* can be obtained as,

$$E\left(x,t\right) = B_1\left|f\left(x,t\right)\right|^2 = \frac{1}{\sqrt{2\pi}} \cdot \frac{\Pi}{\Delta x_0\sqrt{1+\alpha^2}} \exp\left\{-\frac{\left(x-c_1 t\right)^2}{2\Delta x_0{}^2\left(1+\alpha^2\right)}\right\}. \tag{19}$$

A corresponding plot of the energy distribution in *x*-space is presented in Fig. 2 with same parameters. The energy distribution in *q*-space (Eq. (5)) is only determined by the initial perturbation and not varying with time. All solutions are exact up to this stage. On the other hand, we are also able to obtain asymptotic solutions from which we can introduce the concept of locally dominant phonon modes *k*.

## 3. Asymptotic solutions using stationary phase approximation

Equation (10) is an exact integral of Eq. (9). However, we can also apply the stationary phase approximation [21] to obtain the asymptotic solution of integral of Eq. (9) with $t \to \infty$. The phase term in Eq. (9)

$$\Psi\left(q\right) = qx - \omega\left(q\right)t \tag{20}$$

is considered to be stationary only if the phase $\Psi$ satisfies

$$\frac{d\Psi}{dq} = x - \frac{d\omega}{dq}t = 0. \tag{21}$$

The essential idea of stationary phase approximation is the cancellation of sinusoids with rapidly varying phase $\Psi$. By substituting the dispersion relation (Eq. (6)) into Eq. (21), we can identify a position and time dependent wavenumber *k* for stationary phase where

$$k(x,t) = q\big|_{d\Psi/dq=0} = (x - c_1 t)/(2c_2 t). \qquad (22)$$

This wavenumber *k* (different from *q*) has a physical meaning of dominant wavenumber at given location *x* and time *t* (locally dominant phone mode). $\Psi(q)$ can be expanded around *k* using Taylor series,

$$\Psi(q) \approx kx - \omega(k)t - \frac{t}{2}\frac{d^2\omega}{dq^2}\bigg|_{q=k}(q-k)^2. \qquad (23)$$

Substitution of expression (23) back to the original integral Eq. (9) leads to the asymptotic solution of *f(x,t)* ,

$$\lim_{t\to\infty} f(x,t) = \Delta x_0 A e^{-(\Delta x_0 k)^2}\sqrt{2\Big/\left[t\left(d^2\omega/dq^2\right)\big|_{q=k}\right]}\, e^{i(kx-\omega t-\pi/4)}. \qquad (24)$$

The asymptotic solution of integration can be obtained using dispersion relation (Eq.(6)),

$$\lim_{t\to\infty} f(x,t) = \frac{\Delta x_0 A}{\sqrt{c_2 t}} e^{-(\Delta x_0 k)^2} e^{i(kx-\omega t-\pi/4)}, \qquad (25)$$

which can also be recovered from the exact solution (11) with $\alpha = c_2 t/\Delta x_0^2 \to \infty$. The small $\Delta x_0$ and large $c_2$ lead to fast converge of the asymptotic solution (Eq.(25)) to the exact solution (Eq. (11)).

## 4. Locally dominant phonon mode *k* and energy redistribution in *k*-space

The locally dominant phonon mode *k*(*x*,*t*) defined in the stationary phase approximation that is varying with position and time has the physical meaning of dominant vibration mode at given position *x* and time *t*. Generally speaking, it describes the local periodicity. Obviously, it is different from the wavenumber *q* that is a global variable independent of position *x* and time *t*.

$k(x,t)$ is defined locally and is critical to the establishment of local thermodynamic equilibrium (LTE) that requires energy to be uniformly distributed in $k$-space locally.

The asymptotic dependence of locally dominant phonon mode $k$ on $x$ and $t$ was given by Eq. (22). However, the exact temporal and spatial dependence of $k$ can be obtained for this particular second order dispersive chain. The dispersion relation (Eq. (6)) should be exactly satisfied at any position $x$ and time $t$. Therefore, particles in the dispersive chain with locally dominant wavenumber $k$ should have a vibration frequency given by $\omega_a(k) = c_1 k + c_2 k^2$, which is also a position- and time-dependent variable. Hence, the phase term in Eq. (11) can be written as

$$-\frac{D_2}{8}(x - c_1 t)^2 = k(x,t)x - \omega_a(k)t = kx - c_1 kt - c_2 k^2 t. \qquad (26)$$

Substitution of expression of $D_2$ (Eq. (13)) into Eq. (26) leads to the exact solution of $k(x,t)$,

$$k(x,t) = \beta_1 \cdot \frac{(x - c_1 t)}{2c_2 t}, \qquad (27)$$

where the dimensionless number $\beta_1 = \left(1 - 1\big/\sqrt{1+\alpha^2}\right)$ ranges between 0 and 1. Recall that $\alpha = c_2 t \big/ \Delta x_0^2$, the asymptotic solution of $k$ (Eq. (22)) can be recovered from Eq. (27) with time $t \to \infty$ (or $\beta_1 \to 1$). Equation (27) describes the exact variation of locally dominant phonon mode $k$ with position $x$ and time $t$. Figure 3 shows the variation of $k$ with time at three different locations x = -5, 0, and 5 with the solid line for the exact solution (Eq. (27)) and the dash line for the asymptotic solution (Eq. (22)). Convergence of asymptotic solution to exact solution can be obtained for large $t$. It was found that dominant phone mode $k(x, t=0) = 0$ for everywhere because there is no motion at the moment $t$=0. Dominant phonon mode $k(x = c_1 t, t)$ right at the plane wave front is always zero. For any location $x$, the dominant phonon mode $k$ increases in a very short time, and reaches maximum value followed by the slow decrease to the asymptotic value of $-c_1/2c_2$ .

We have derived the energy distribution in wavenumber *q*-space (Eq. (5)) and in position *x*-space (Eq. (19)). It was shown that the energy distribution in wavenumber *q*-space is stationary for linear problem, even with a nonlinear dispersion. Recall that *k* is locally dominant phonon mode that is different from the global wavenumber *q*, it will be very interesting to investigate the energy distribution in *k*-space. Temporally and spatially varying *k(x,t)* leads to the energy redistribution among the phonon modes in *k*-space. The energy distribution in *k*-space can be derived using the chain rule,

$$E(k,t)=\frac{E(x,t)}{dk/dx}=\Pi\sqrt{\frac{2}{\pi}}\frac{\Delta x_0}{\beta_2}e^{-2k^2\Delta x_0^2/\beta_2^2}\ , \qquad (28)$$

where the dimensionless number $\beta_2=\left(\sqrt{1+\alpha^2}-1\right)/\alpha$ represents the spatial spread of energy distribution in *k*-space. A plot of both $\beta_1$ in Eq. (27) and $\beta_2$ in Eq. (28) is presented in Fig. 4, both of which range between 0 and 1, and monotonically increase with $\alpha$ or *t*. The energy redistribution in *k*-space is plotted in Fig. 5. At the very beginning, energy distribution in *k*-space is nothing but a delta function where all energy concentrates in the only relevant mode, namely the long wave mode *k*=0. Energy starts to redistribute in *k*-space with time *t* and eventually approaches a steady-state distribution, the same distribution as the energy distribution in *q*-space (Eq. (5)). The time of transition to the steady state distribution in *k*-space is dependent on the parameter $c_2$ and the width of initial displacement $\Delta x_0$. A large $c_2$ or small $\Delta x_0$ leads to a fast transition to the steady state energy distribution, which is in agreement with our intuition that perturbation with smaller $\Delta x_0$ tends to dissipate faster.

It was found that the energy redistribution in *k*-space is only possible for dispersive chain ($c_2 \neq 0$), and this distribution will eventually approach a steady state distribution. A timescale $\lambda = \Delta x_0^2 / c_2$ can be identified with this redistribution process. The initial perturbation will never excite any locally dominant phonon modes in non-dispersive chain with linear dispersion ($c_2 = 0$), where the energy redistribution in *k*-space is strictly prohibited and the energy redistribution in *k*-space can never reach the stead-state distribution (i.e. the distribution in *q*-space). Especially for unit perturbation ($\Delta x_0 \to 0$) that drives system away from equilibrium, energy is uniformly distributed in *q*-space. However, the transition from this nonequilibrium state to steady state energy distribution in *q*-space does not happen instantaneously. It requires finite time that is related to the time scale $\lambda$. Particularly, for this case the energy distribution in *k*-space for non-dispersive chain can be a constant dirac-delta function. This transition to the steady state uniform distribution only happens in dispersive chain.

From the microscopic view, a perturbation (mechanical energy) introduced into a system initially in equilibrium will drive that system away from equilibrium. First, the energy associated with the perturbation will be redistributed into the local phonon modes (phonon excitation) in *k*-space due to the nonlinear dispersion. The interactions of excited phonons (from the initial perturbation) with thermal phonons (from thermal vibration) at the same location and/or impurity (defect) via other mechanisms will eventually thermalize the system and restore the system back to the initial equilibrium state. For perturbation with smaller $\Delta x_0$, for example the displacement of a single atom, the quick redistribution (short timescale $\lambda$) and uniform energy distribution in *k*-space (in all phonon modes) enables a faster establishment of a new thermodynamic equilibrium state. The perturbation with larger $\Delta x_0$ (coordinated atomic motion with long wave length) has a much longer redistribution timescale $\lambda$. Most energy is concentrated in the long wave phonons ( $k$ close to 0) and this leads to a much slower relaxation to the new thermodynamic equilibrium state.

## 5. One-dimensional harmonic chain

Discrete system is inherently dispersive. One-dimensional harmonic chain (as shown in Fig. 6) is one of the most studied classical models. The corresponding dispersion relation reads

$$\omega = 2\sqrt{\frac{\chi}{m}}\sin\left(\frac{qa}{2}\right) \tag{29}$$

for harmonic chain with nearest neighboring interaction, where $m, \chi, a$ are the mass, spring constant and lattice constant. No close-form exact analytical solutions can be found for an initial perturbation in the form of Eq. (2). However, asymptotic solutions are available using the stationary phase approximation. First, we rewrite the dispersion relation as

$$\omega = \frac{2}{a}c_1\sin\left(\frac{qa}{2}\right). \tag{30}$$

An approximation via Taylor expansion up to the third order can be written as

$$\omega \approx c_1 q + c_2 q^2 + c_3 q^3, \tag{31}$$

where

$$c_1 = a\sqrt{\frac{\chi}{m}}, \quad c_2 = 0, \quad c_3 = -\frac{a^3}{24}\sqrt{\frac{\chi}{m}} = -\frac{a^2}{24}c_1. \tag{32}$$

Similar to the example with second order dispersion in Section 2, the wave propagating with dispersion (31) satisfies the partial differential equations [17-19]

$$u_{,t} + c_1 u_{,x} = c_3 u_{,xxx}, \tag{33}$$

or

$$u_{,tt} = c_1^2 u_{,xx} - 2c_1 c_3 u_{,xxxx} \,. \tag{34}$$

Eq. (33) is also known as the linearized KdV equation. By substituting the dispersion relation (Eq.(30)) into the stationary phase Eq. (21), we can identify the locally dominant phonon wavenumber $k$,

$$k(x,t) = q\big|_{d\Psi/dq=0} = \pm k_0 = \pm a\cos\left(\frac{x}{c_1 t}\right)\cdot\frac{2}{a} \tag{35}$$

for $|x| < c_1 t$ , $|k| < \pi/a$ , where $\pi/a$ is the maximum mode of first Brillouin zone. Recall that relation (35) is an asymptotic solution with $t \to \infty$ (not like Eq. (27) which is exact), a plot of *k(x,t)* variation with *x* and *t* is shown in Fig. 7 indicating that local dominant phonon mode *k(x,t)* increases from zero to $\pi/a$ with time, similar to the example in Section 2 with second order nonlinear dispersion relation. The corresponding frequency at given location and time is

$$\omega_a(x,t) = \frac{2}{a} c_1 \sqrt{1 - \frac{x^2}{(c_1 t)^2}} \,. \tag{36}$$

Now we introduce a displacement perturbation given by Eq. (2) to an initially static harmonic chain. The dynamics can be obtained by decompose the displacement fields $u$ into the wave solutions traveling in the positive direction ( $f^+$ ) and the negative direction ( $f^-$ ), where $u = f^+ + f^-$ .

$$f^+(x,t) = \frac{\Delta x_0 A}{2\sqrt{\pi}} \int_{-\infty}^{\infty} e^{-(\Delta x_0)^2 q^2} e^{i(qx - \omega t)} dq \,, \tag{37}$$

where the amplitude is $A/2$. The integral of Eq. (37) can be obtained using the stationary phase method with the dispersion relation in Eq. (30),

$$f^{+}(x,t)=\frac{B}{\sqrt{t\left|d^{2}\omega/dq^{2}\right|_{q=k}}}\frac{\Delta x_{0}A}{2\sqrt{\pi}}e^{-(\Delta x_{0}k)^{2}}e^{i\left(kx-\omega_{a}t+\frac{\pi}{4}\right)},\tag{38}$$

where $B$ is a scaling factor that can be computed through energy normalization and

$$\left.\frac{d^{2}\omega}{dq^{2}}\right|_{q=k}=-\frac{a}{2}c_{1}\sin\left(\frac{ka}{2}\right)=-\frac{1}{4}a^{2}\omega_{a}\tag{39}$$

according to dispersion relation in Eq. (30). Finally,

$$f^{+}(x,t)=\frac{B}{a\sqrt{t\omega_{a}}}\frac{\Delta x_{0}A}{\sqrt{\pi}}e^{-(\Delta x_{0}k)^{2}}e^{i\left(kx-\omega_{a}t+\frac{\pi}{4}\right)}.\tag{40}$$

The energy associated with $f^{+}$ is

$$E(x,t)=B_{1}\left|f^{+}\right|^{2}=\frac{\Pi}{\sqrt{2\pi}}\frac{B^{2}}{a^{2}\omega_{a}t}\frac{\Delta x_{0}}{\pi}e^{-2(\Delta x_{0}k)^{2}}.\tag{41}$$

The energy distribution in $k$-space is

$$E(k,t)=E(x,t)\frac{dx}{dk}=\frac{\Pi}{\sqrt{2\pi}}\frac{B^{2}\Delta x_{0}}{4\pi}e^{-2(\Delta x_{0}k)^{2}}.\tag{42}$$

Since the total energy should be conserved,

$$\int_{-\pi/a}^{\pi/a}E(k,t)dk=\Pi,\tag{43}$$

the normalization constant $B$ can be obtained as

$$B=\frac{2\sqrt{2\pi}}{\sqrt{erf\left(\sqrt{2}\pi\Delta x_{0}/a\right)}}.\tag{44}$$

Finally, the fundamental solution of $u$ with given perturbation can be written as,

$$u(x,t)=\frac{4\sqrt{2}}{\sqrt{erf\left(\sqrt{2}\pi\Delta x_{0}/a\right)}}\cdot\frac{\Delta x_{0}A}{a\sqrt{t\omega_{a}}}e^{-(\Delta x_{0}k)^{2}}\cos\left(kx-\omega_{a}t+\frac{\pi}{4}\right).\tag{45}$$

Equation (45) can be used to find the elementary solution (Green's function) for unit perturbation in the harmonic chain with $\Delta x_0 \ll a$ . A comparison between the molecular dynamics simulation of one-dimensional harmonic chain and asymptotic solution from Eq. (45) showing good agreement is presented in Fig. 8, where parameters $A = \Delta x_0 = 0.1$ $c_1 = a = 1$ are used. The energy distribution in *x*-space can be obtained using Eq. (19), where

$$E\left(x,t\right) = E(k,t)\frac{dk}{dx} = \frac{\Pi}{\sqrt{\pi}}\frac{4\sqrt{2}}{erf\left(\sqrt{2}\pi\Delta x_0/a\right)}\frac{\Delta x_0}{a^2\omega_a t}e^{-2(\Delta x_0 k)^2} . \quad (46)$$

We need to point out that the expression $dk/dx$ has a physical meaning of phonon density. Plot of the energy distribution in *x*-space (Eq. (46)) for harmonic chain is presented in Fig. 9 for two different $\Delta x_0$ . Energy is more concentrated in space right before the wave front with higher density of phonons. It is more concentrated in local space for larger $\Delta x_0$ , which is expected for a more widely spread perturbation. Also it is found that energy distribution is decaying as $t^{-1}$ for constant *k* or $x/(c_1 t)$ at wave front that is the same as second order nonlinear dispersion (Eq. (19) with $t \to \infty$ ). The energy distribution in *k*-space is

$$E\left(k,t\right) = \frac{\Pi}{\sqrt{\pi}}\frac{\sqrt{2}\Delta x_0}{erf\left(\sqrt{2}\pi\Delta x_0/a\right)}e^{-2(\Delta x_0 k)^2} . \quad (47)$$

Obviously the energy distribution in *k*-space is the asymptotic solution with $t \to \infty$ and therefore no time varying solution like the exact solution (Eq. (28)) can be obtained for harmonic chain. It is also confirmed that the limiting steady state energy distribution in *k*-space with $t \to \infty$ does not dependent on the dispersion relation as the distributions for nonlinear second order dispersion (Eq. (28) and Eq. (47)) are essentially the same, which is also expected. However, the transition to the steady state distribution in *k*-space should dependent on the particular dispersion.

Solutions can also be obtained for the approximate dispersion (Eq. (31)) that is exact for linearized KdV equation via the same procedure and will not repeat here.

## 6. Heat transport along the harmonic chain and local thermodynamic equilibrium

In this section, we will apply the previous results to study the heat transport along the harmonic chain that is initially stationary. The example used here is shown in Fig. 6, where all atoms $\leq n$ (on the left side) was given a prescribed temperature of $T_0$ at time t=0. The thermal vibration of the atoms with temperature $T_0$ mimics the perturbations introduced into the atomic chain on all atoms $\leq n$ that leads to the continuously wave propagating to the right part which is initially stationary, and the energy starts to transport along the atomic chain. Obviously this can be analyzed using the principle of superimposition of our elementary solutions derived in Section 6 for unit perturbation with $\Delta x_0 \ll a$, which is appropriate to represent the thermal vibration.

At any given location *x*, the energy transport leads to the continuous excitation of dominant phonons starting from $k=0$ at time $t = x/c_1$ to the maximum $k = k_{\max}$ with increasing time *t*, where

$$k_{\max}(x,t) = a\cos\left(\frac{x}{c_1 t}\right)\cdot\frac{2}{a}. \tag{48}$$

With increasing time *t*, $k_{\max}$ at that location will eventually approach the value $\pi/a$, namely the maximum mode in the first Brillouin zone. The energy distribution among various phonon modes at the same location should be uniform for small $\Delta x_0 \ll a$. However, local thermodynamic equilibrium (LTE) at given location *x* cannot be established within finite time without the phonon scattering mechanism that is missing in the harmonic chain. The fact $k_{\max} < \pi/a$ indicates that energy cannot be uniformly distributed in the entire first Brillouin zone at any position.

The energy distribution in *x*-space is nothing but proportional to the number of phonon modes at that location, such that

$$E(x,t) = \Pi^* \frac{k_{\max}}{\pi/a} = \Pi^* \cdot a\cos\left(\frac{x}{c_1 t}\right) \cdot \frac{2}{\pi}, \tag{49}$$

where $\Pi^* = T_0 C_v$ is the energy density in the left part of the chain and $C_v$ is the heat capacity. The energy flux can be computed as

$$J = \frac{d\int_0^{c_1 t} E(x,t)\,dx}{dt} = \Pi^* \frac{2}{\pi} c_1. \tag{50}$$

Obviously this ballistic energy transport is at the sound speed $c_1$ with constant energy flux, which is consistent with the findings from previous study on the harmonic chains [7]. The same calculation via Fourier law leads to the temperature solution,

$$T = T_0 \left\{1 - erf\left(\frac{x}{\sqrt{4\kappa t/C_v}}\right)\right\}, \tag{51}$$

where $\kappa$ is the thermal conductivity. The energy flux can be computed as,

$$J = -\kappa \frac{dT}{dx} = \frac{\Pi^*}{\sqrt{\pi}} \sqrt{\frac{\kappa}{C_v t}} \tag{52}$$

By equating the two energy fluxes (Eqs. (52) and (50)), an equivalent thermal conductivity can be obtained as

$$\kappa = 4c_1^2 C_v t/\pi \tag{53}$$

that is proportional to heat capacity $C_v$ but diverges with time *t*.

## 7. Conclusions

Fundamental solutions of wave propagation in one-dimensional dispersive chain subject to a localized initial perturbation are presented for both second order nonlinear dispersive chain and homogenous harmonic chain. Solutions are compared to the molecular dynamics simulations and good agreement can be obtained. The locally dominant phonon modes $k(x,t)$ are introduced which is different from the global wavenumber $q$. The variations of local modes with time and space are critical to the energy redistribution in $k$-space and the establishment of local thermodynamic equilibrium (LTE).

The energy redistribution among various phonons modes ($k$-space) is only possible with nonlinear dispersion and requires a finite amount of time to reach the limiting steady state distribution that is the same as energy distribution in $q$ space. The transition time is dependent on the spatial distribution of initial perturbation and the dispersion relation. Energy transport along the harmonic chain will excite the local phonon modes starting from the lowest mode ($k$=0) and gradually expanding to the highest mode ( $k = k_{\max}(x,t)$ ) with increasing time. Energy is uniformly distributed in all available local phonon modes, but not the entire first Brillouin zone. The LTE cannot be fully established in harmonic chain as the maximum mode $k_{\max}(x,t)$ can only asymptotically approach the upper limit of first Brillouin zone. The energy flux along the chain is a constant with time and proportional to the sound speed (ballistic transport). A time-dependent thermal conductivity can be obtained and this conductivity diverges with time.

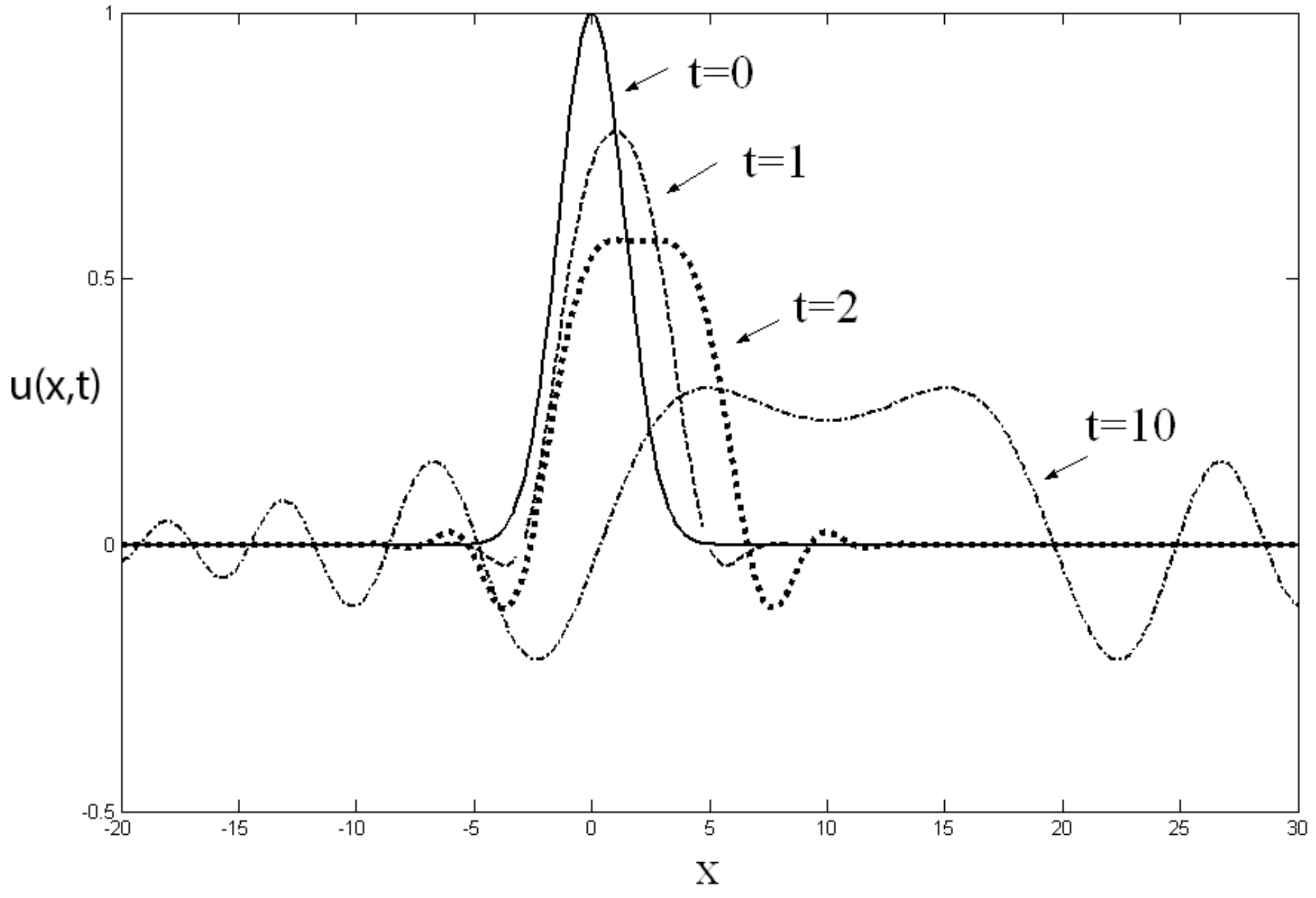

Figure 1. Time evolution of displacement u(x,t) along *x*+ direction

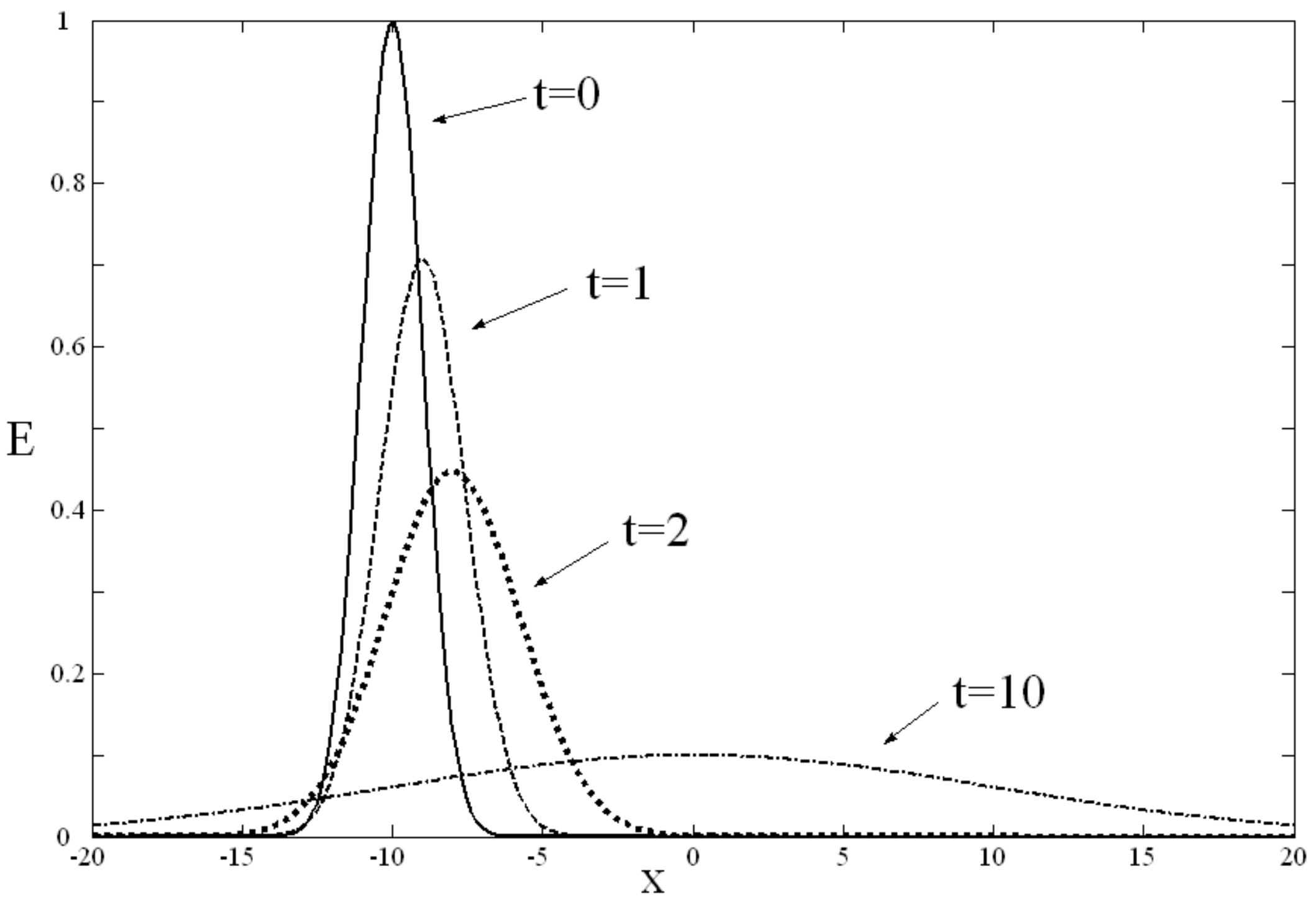

Figure 2. Time evolving of energy distribution in *x*-space.

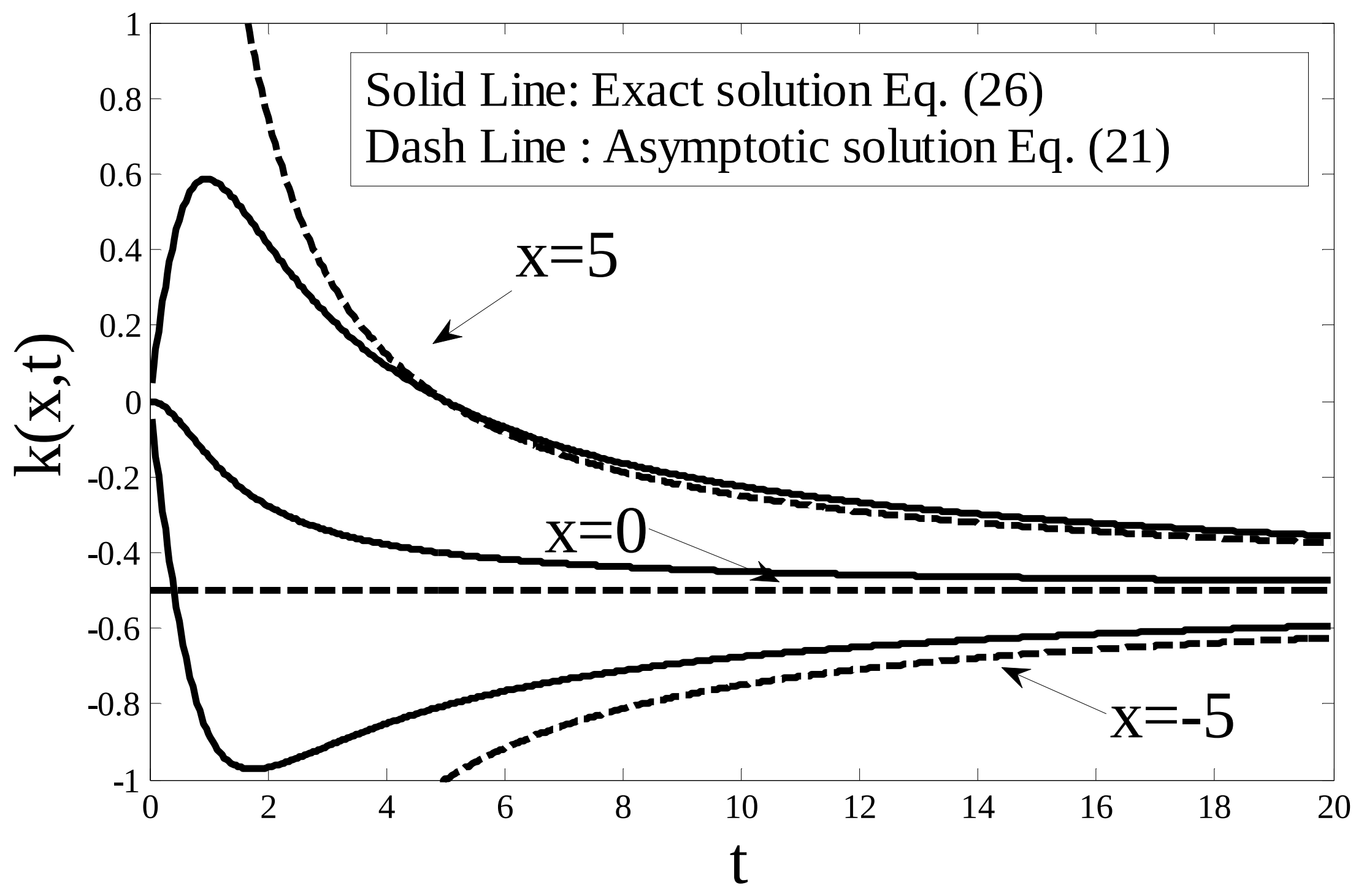

Figure 3. Time evolving of locally dominant phonon mode *k(x,t)* at different locations. The solid line shows the exact solution and the dash line denotes the asymptotic solution.

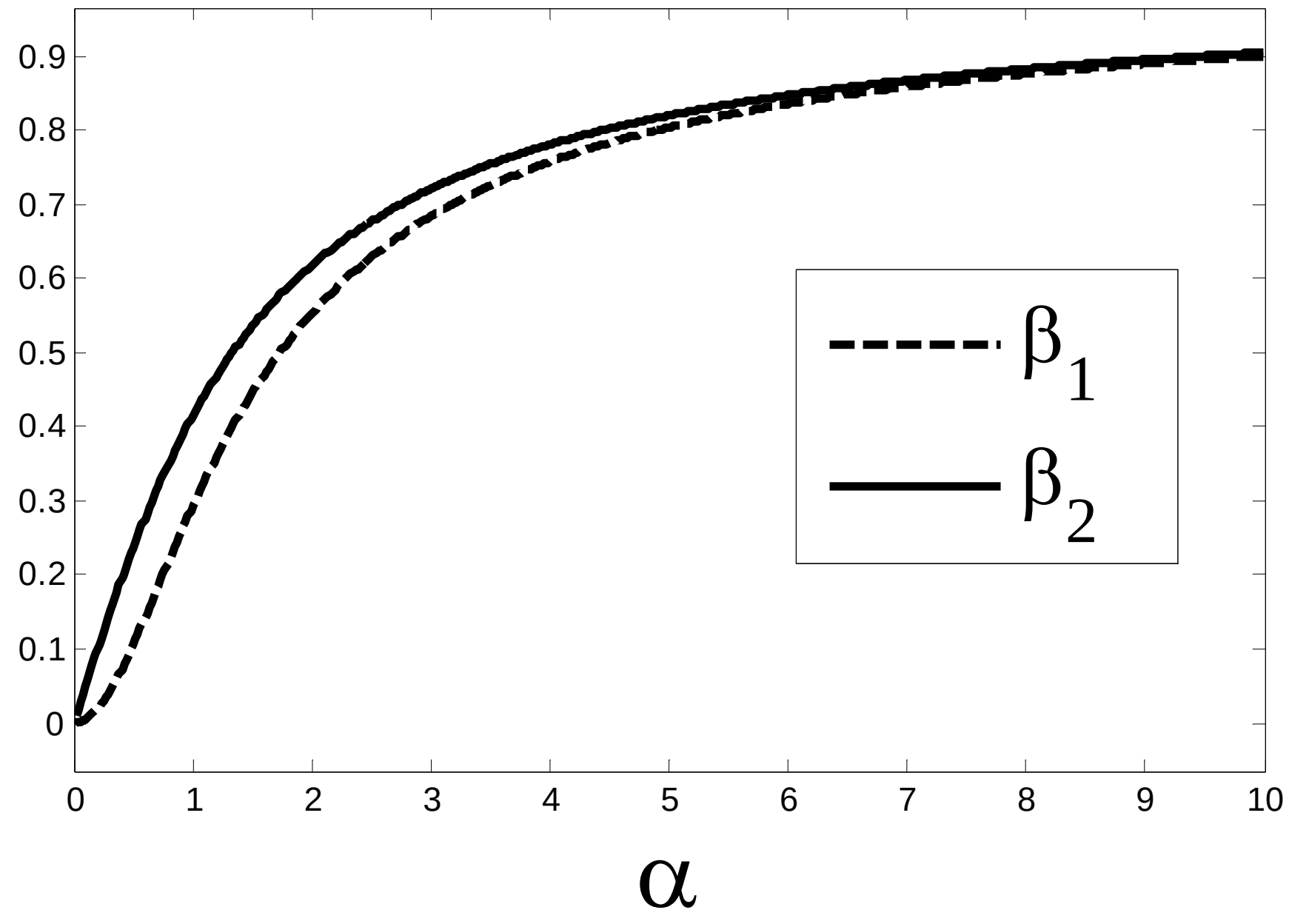

Figure 4. Variation of two dimensionless numbers $\beta_1$ and $\beta_2$ with $\alpha$ .

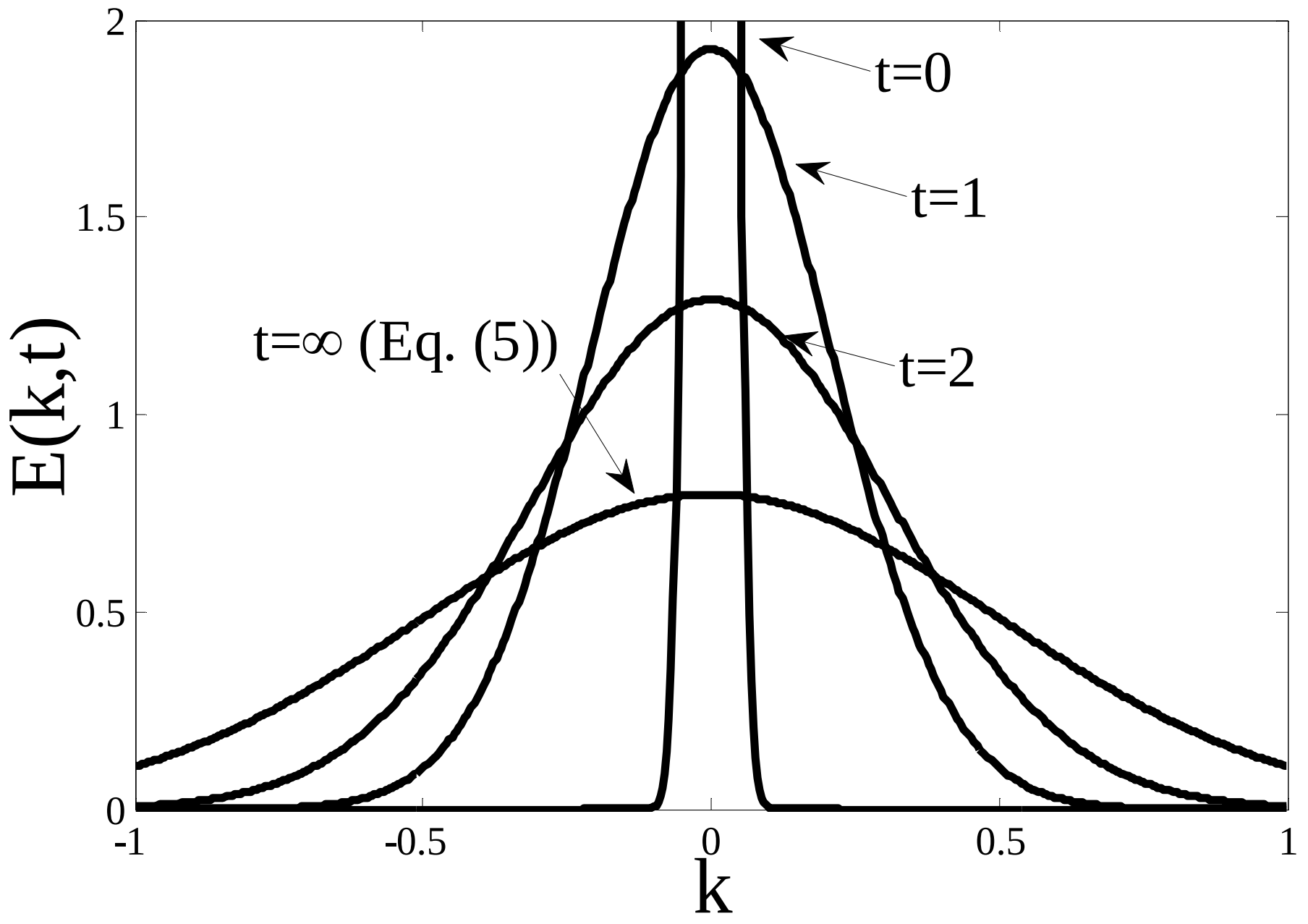

Figure 5. Time evolving of energy distribution in *k*-space among various phonon modes.

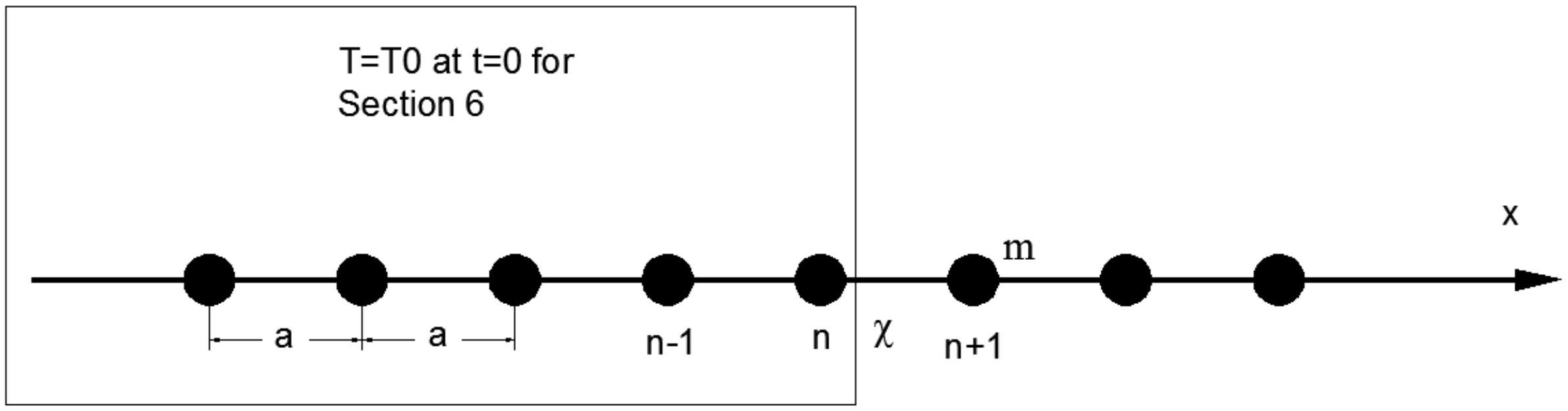

Figure 6. Schematic plot of one-dimensional harmonic chain used in Section 5 and 6. In Section 6, all atoms with number $\leq n$ are given a prescribe temperature $T_0$ at time t=0.

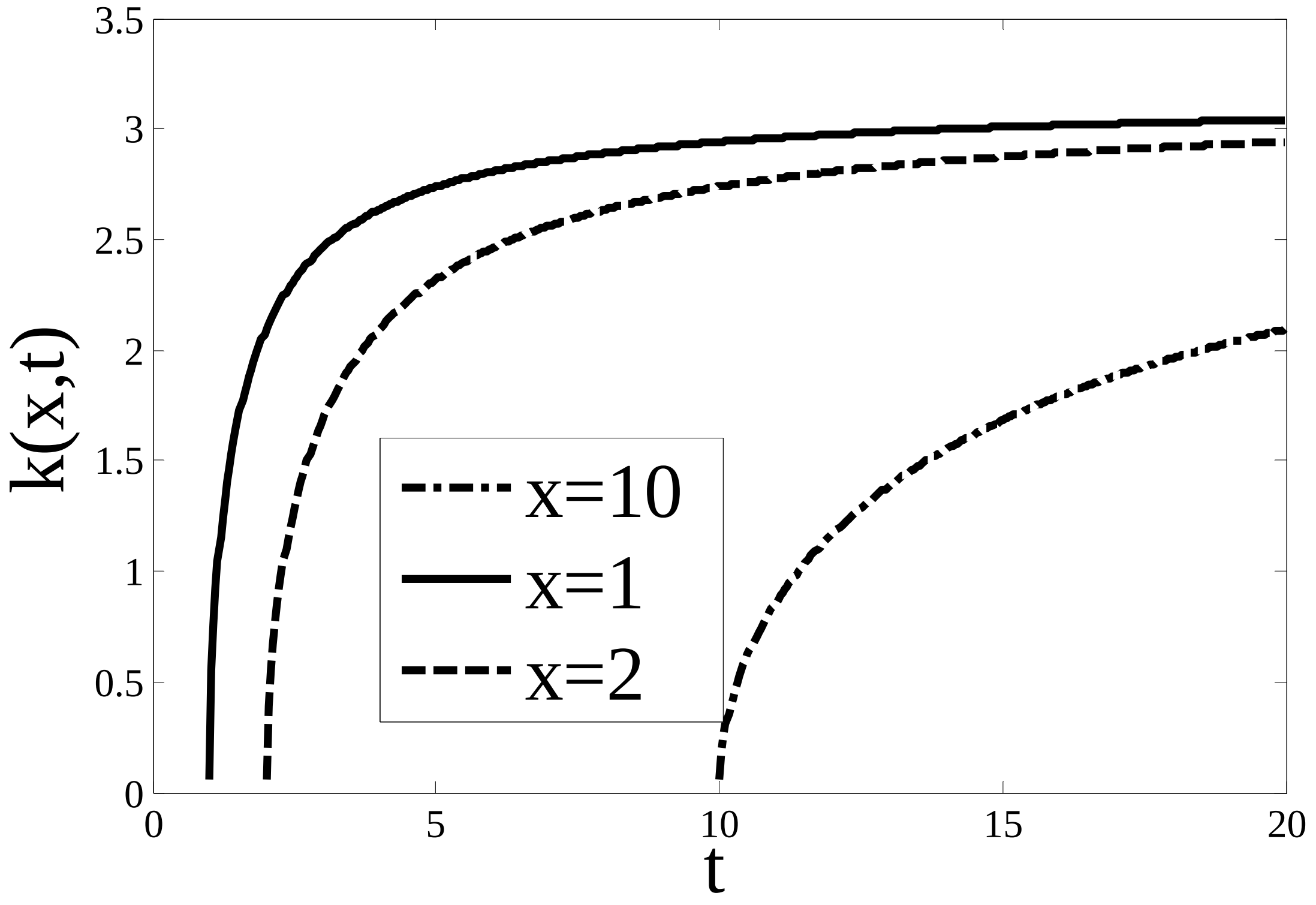

Figure 7. Time evolving of locally dominant phonon modes *k(x,t)* for harmonic chain.

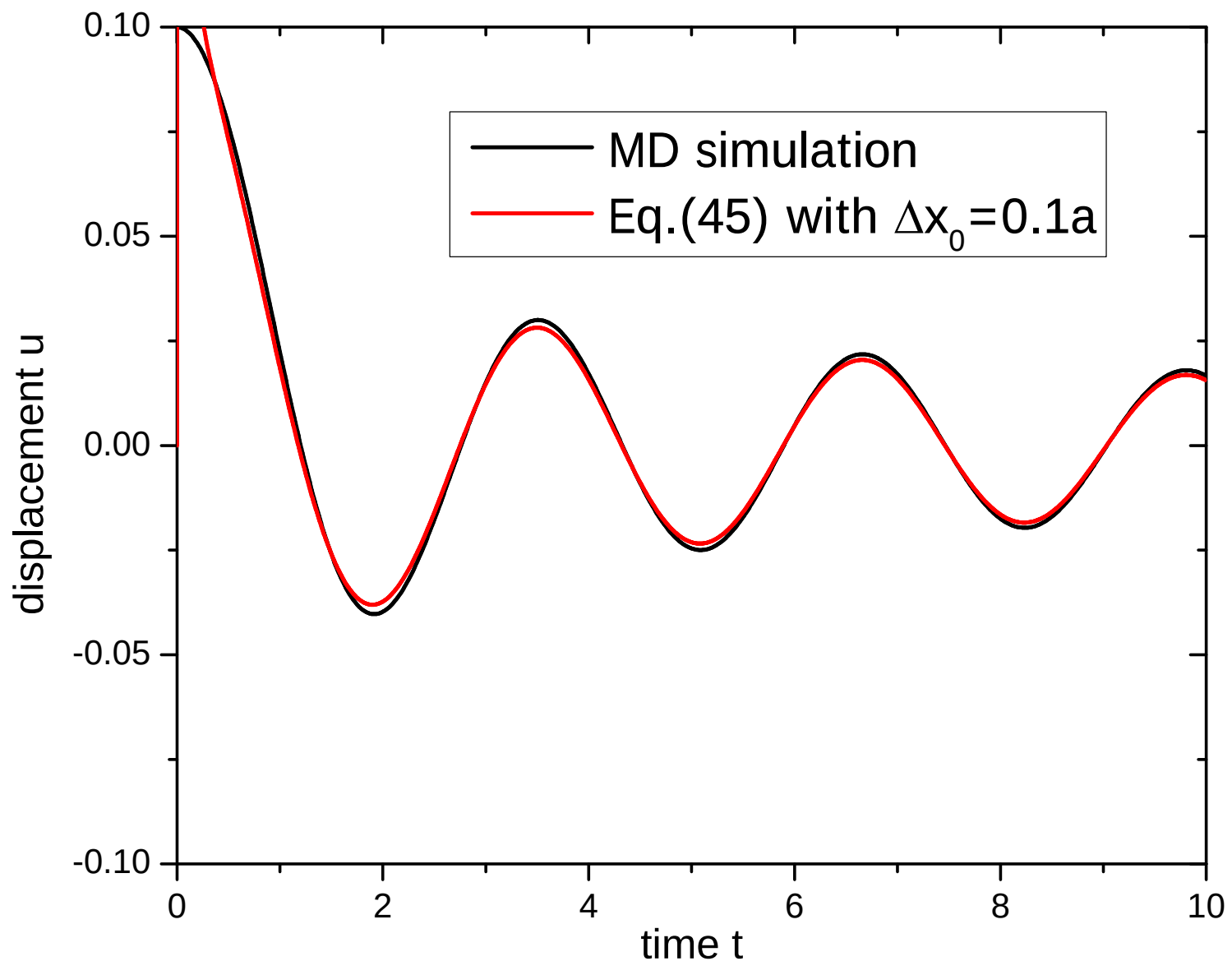

Figure 8. The time evolution of displacement *u(x=0,t)* with initial displacement perturbation. Good agreement is obtained for results from molecular dynamics simulations and from the analytical solution Eq. (45).

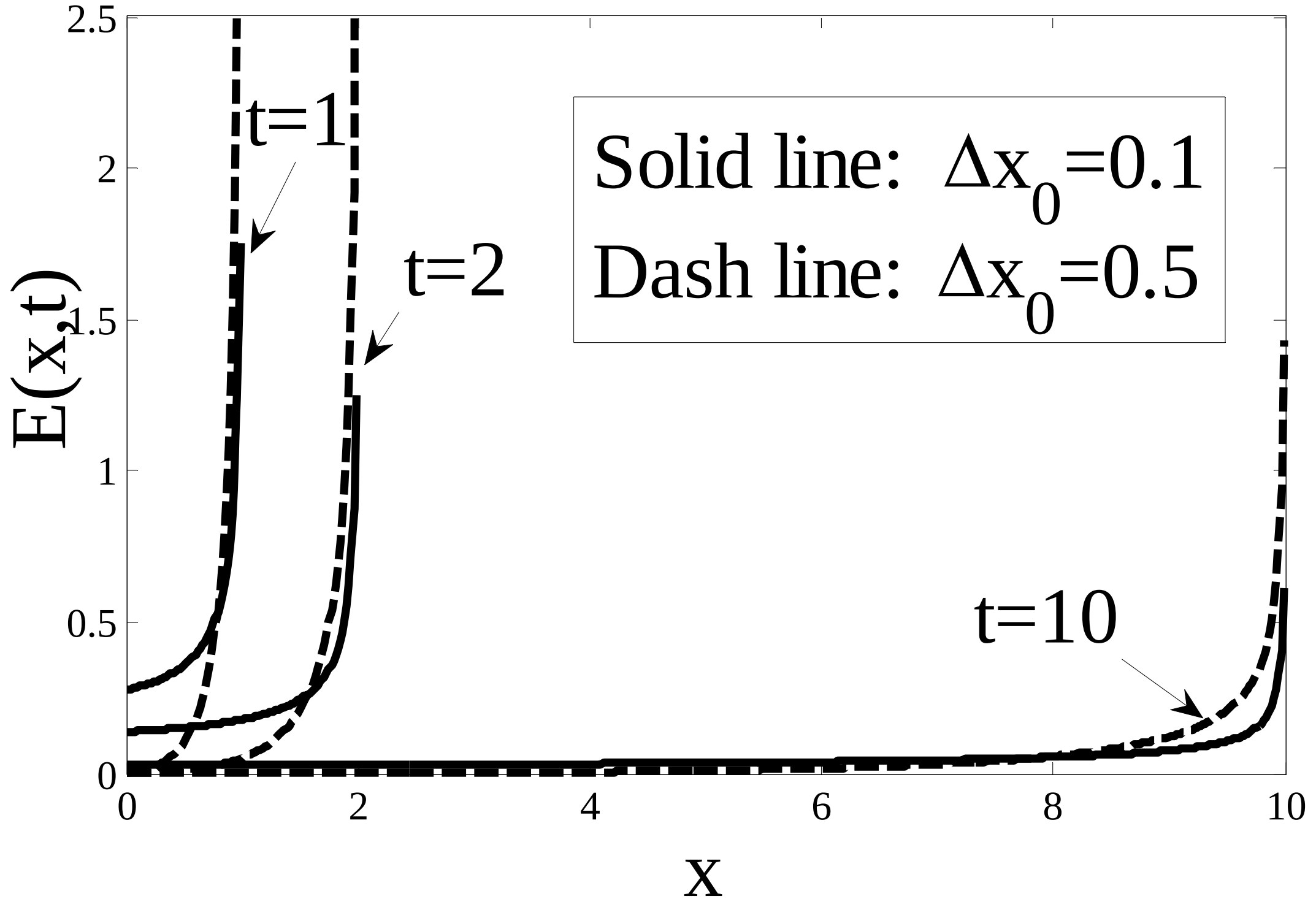

Figure 9. Energy distribution in *x*-space at various time for harmonic chain. The solid line with smaller $\Delta x_0$ shows a more spread, wider, and faster decay of energy distribution.

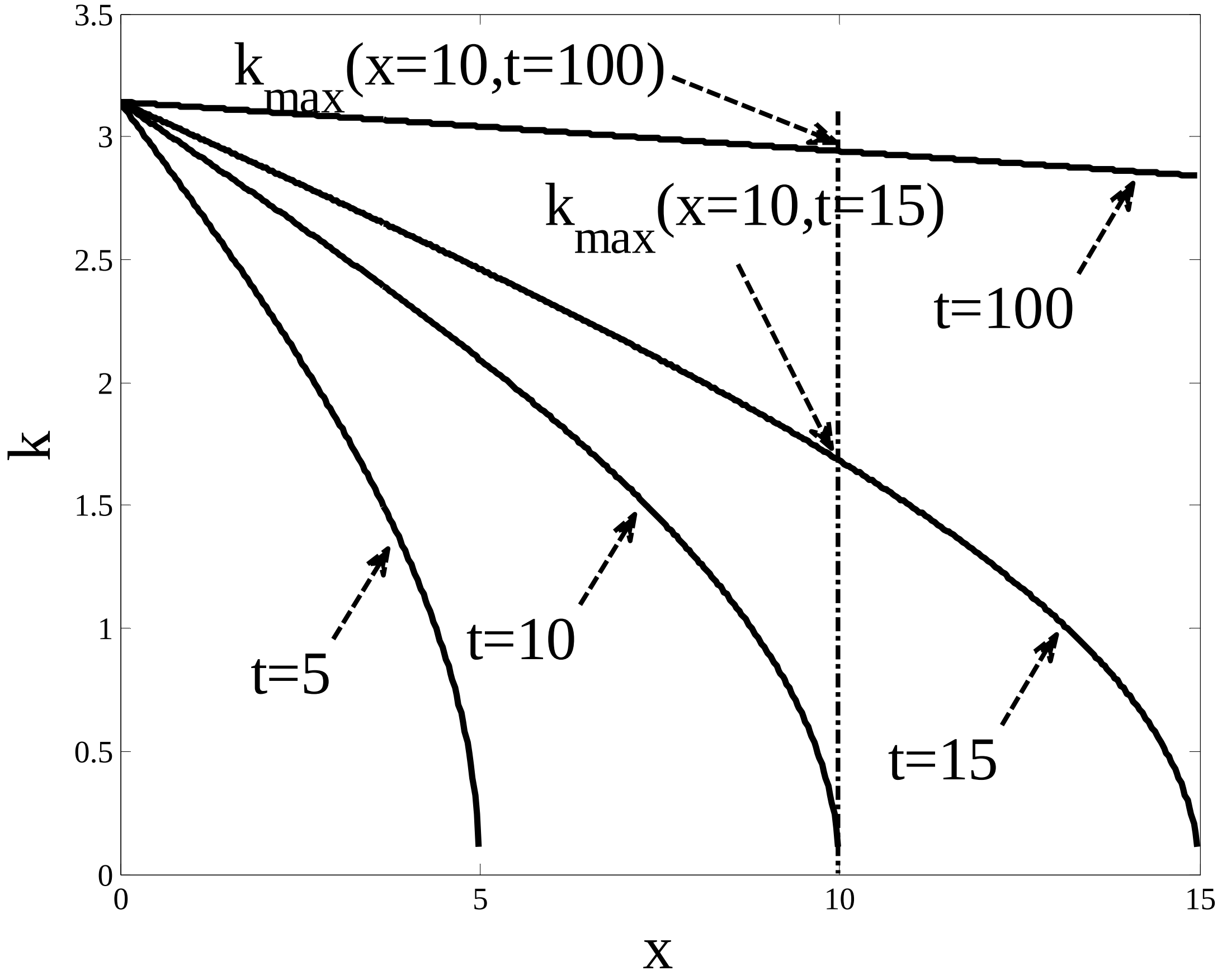

Figure 10. The excitation of local phonon modes with increasing time for heat transport in Section 6. Plot shows the local phonon modes get excited starting from zero to $k_{\max}$ with energy uniformly distributed in all available modes $k \le k_{\max}$. However, the local thermodynamic equilibrium (LTE) cannot be established with $k_{\max} < \pi/a$.